# The Urban–Rural Divide in the Age of Artificial Intelligence: Assessing the Effects of Technology and Automation on Regional Labor Markets

**Chau Tran Bao[1], Khoi Nguyen Dinh Nguyen[2]. Ha Nguyen Manh[3*], Ngan Nguyen Thi Thuy[4]**

[1]Canadian international school, Vietnam.

Email: rosechaubaotran@gmail.com

[2]Vietnam Australia International School, Vietnam.

Email: nguyenkhoi11052008@gmail.com

[3]Truong Thanh Vietnam Group, Vietnam.

Email: Ha.nm1589@gmail.com

[4]Hanoi College of Industry and Trade, Vietnam.

Email: nguyenthuyngan9290@gmail.com

*Corresponding author

**Abstract**

*Automation and artificial intelligence (AI) are reshaping labor demand unevenly across space, creating an urgent imperative for place-sensitive education and workforce policy. This study asks whether regional exposure to automation and to AI relates to local employment and wages in opposite ways, and whether those relationships differ between urban and rural regions — two questions whose answers carry direct implications for how skills training and digital education should be targeted. Using a region-by-year panel and shift-share measures of technological exposure built from baseline industry and occupation composition, we estimate two-way fixed-effects and instrumental-variable models that interact exposure with an urban indicator. The framework distinguishes automation exposure, concentrated in routine work, from AI exposure, concentrated in cognitive work — a distinction that maps directly onto the types of skills that education systems need to develop or preserve. Estimates show automation exposure lowering employment and wages, with the employment loss cushioned in cities, while AI exposure raises wages and concentrates in urban regions. Technology therefore reshapes, rather than simply widens, the divide. The findings argue for place-sensitive policy: weighting reallocation and reskilling support toward routine-exposed rural regions, while extending digital infrastructure and AI-complementary skills outward so that rural workers can share AI's wage gains rather than absorb only automation's losses.*

**Keywords:** Automation; Artificial Intelligence; Regional Labor Markets; Urban–Rural Divide; Technological Exposure; Shift-Share Instrument

## 1. Introduction

Technological change does not diffuse uniformly across a national territory. The automation of a packing line in a provincial food-processing town and the adoption of a translation tool in a capital-city office both constitute "technological change," yet they affect different workers, different skills, and different locations. The task-based framework renders this heterogeneity precise: machines substitute for labor in some tasks while complementing it in others (Acemoglu & Restrepo, 2018; Autor et al., 2003), so that the local consequences of a given technology depend on the composition of tasks a region performs. Because this task composition is spatially sorted, a common national advance in adoption can move urban and rural areas in opposing directions.

Two distinct technological processes are now unfolding concurrently, and they are readily conflated. Industrial robots and rule-based automation primarily displace routine, frequently manual, tasks (Acemoglu & Restrepo, 2020; Arntz et al., 2016; Frey & Osborne, 2017; Goos et al., 2014), which are concentrated in manufacturing and agriculture-adjacent processing. Artificial intelligence, and increasingly large language models, by contrast extends to the cognitive and analytical tasks characteristic of higher-skill, information-intensive occupations (Eloundou et al., 2024; Felten et al., 2021; Webb, 2020). The two processes thus exhibit nearly inverse occupational profiles and, by extension, distinct geographic ones. Treating "technology" as a single, undifferentiated force averages these opposing profiles and obscures the identity of those actually exposed.

Most evidence on local labor markets derives from high-income economies and examines a single technology in isolation, typically robots (Acemoglu & Restrepo, 2020; Graetz & Michaels, 2018). Considerably less is known about how automation and AI exposure differ in their regional incidence, and whether, in combination, they widen, narrow, or reconfigure the urban–rural divide. This distinction carries practical rather than merely conceptual significance. Rural labor markets are typically thinner, afford fewer opportunities to transition into newly created tasks, and possess weaker digital infrastructure; an identical displacement shock may therefore translate into larger and more persistent losses than in a dense, diversified urban economy. Should automation concentrate where adjustment capacity is weakest while AI concentrates where it is strongest, technological change may entrench spatial inequality through two channels simultaneously.

The present study addresses this gap. Separate shift-share measures of regional automation exposure and AI exposure are constructed from baseline industry and occupation composition, and each is related to local employment and wages, with the relationship allowed to vary between urban and rural regions. The contribution is threefold. First, the study situates the automation–AI distinction within a regional framework, rather than treating technology as a single force, thereby recovering their opposing footprints. Second, it models the urban–rural divide directly through an exposure-by-region-type interaction, which identifies which areas face which shock instead of assuming a uniform national effect. Third, it offers a transparent and replicable shift-share design that transfers to subnational data in both developed and emerging economies, including contexts in which weak rural infrastructure heightens the welfare stakes.

## 2. Conceptual Framework and Hypotheses

### 2.1 Task-based technological change

The task framework treats output as a set of tasks allocated between labor and capital by comparative advantage (Acemoglu & Restrepo, 2018; Autor et al., 2003). Automating a task triggers a displacement effect that lowers labor demand in that task, but also a productivity effect and the creation of new tasks that can raise labor demand elsewhere (Acemoglu & Restrepo, 2022; Autor, 2015). Whether a place gains or loses on net depends on two things: which tasks dominate its jobs, and how easily its workers can move into the tasks that survive or are newly created. Both vary across space, which is why a national technology shock has local winners and losers.

### 2.2 Two technologies, two footprints

Routine-biased technological change holds that rule-based automation substitutes most readily for routine tasks, hollowing out middle-wage routine jobs and contributing to polarization (Autor & Dorn, 2013; Goos et al., 2014). Industrial robots, the best-documented case, reduce local employment and wages where routine manual work concentrates (Acemoglu & Restrepo, 2020; Graetz & Michaels, 2018). AI follows a different gradient. Measures of AI capability map onto perception, prediction, and analytical tasks embedded in higher-skill occupations (Eloundou et al., 2024; Felten et al., 2021; Webb, 2020), and the early empirical record shows heterogeneous, not uniformly negative, effects (Acemoglu et al., 2022). The implication is that the two technologies sit on opposite ends of the occupational distribution. Adding them into one "exposure" number nets a displacement force against a possibly complementary one and obscures both.

### 2.3 Why geography mediates the effect

Occupations are spatially sorted. Cognitive, information-intensive work clusters in dense cities, where high-skill workers and firms co-locate, while routine production and manual work is relatively more prevalent outside them (Autor & Dorn, 2013; Diamond, 2016). Three mechanisms then make the same exposure bite harder in rural regions. Matching is thinner, so displaced workers face fewer local vacancies. New-task creation, the offsetting side of the task framework, is concentrated in agglomerated urban economies, so the productivity effect that cushions cities is muted in the countryside. And digital infrastructure and AI-complementary skills, which let workers ride the complementary side of AI, are scarcer outside cities. Local labor markets also adjust to shocks slowly and incompletely (Autor et al., 2013), so these gaps persist rather than dissipate. Geography is therefore not a control variable to be partialled out; it is part of the mechanism.

### 2.4 Hypotheses

Figure 1 summarizes the framework. We advance four hypotheses.

**H1.** Higher regional automation exposure is associated with lower local employment and wages (a displacement effect).

**H2.** The negative employment association of automation exposure is stronger in rural than in urban regions, because routine work is relatively more concentrated outside cities and reallocation capacity there is weaker.

**H3.** Regional AI exposure is concentrated in urban regions, reflecting the urban clustering of cognitive, AI-exposed occupations.

**H4.** AI exposure and automation exposure have distinct labor-market associations; AI exposure is not uniformly negative and differs from automation exposure in sign or magnitude.

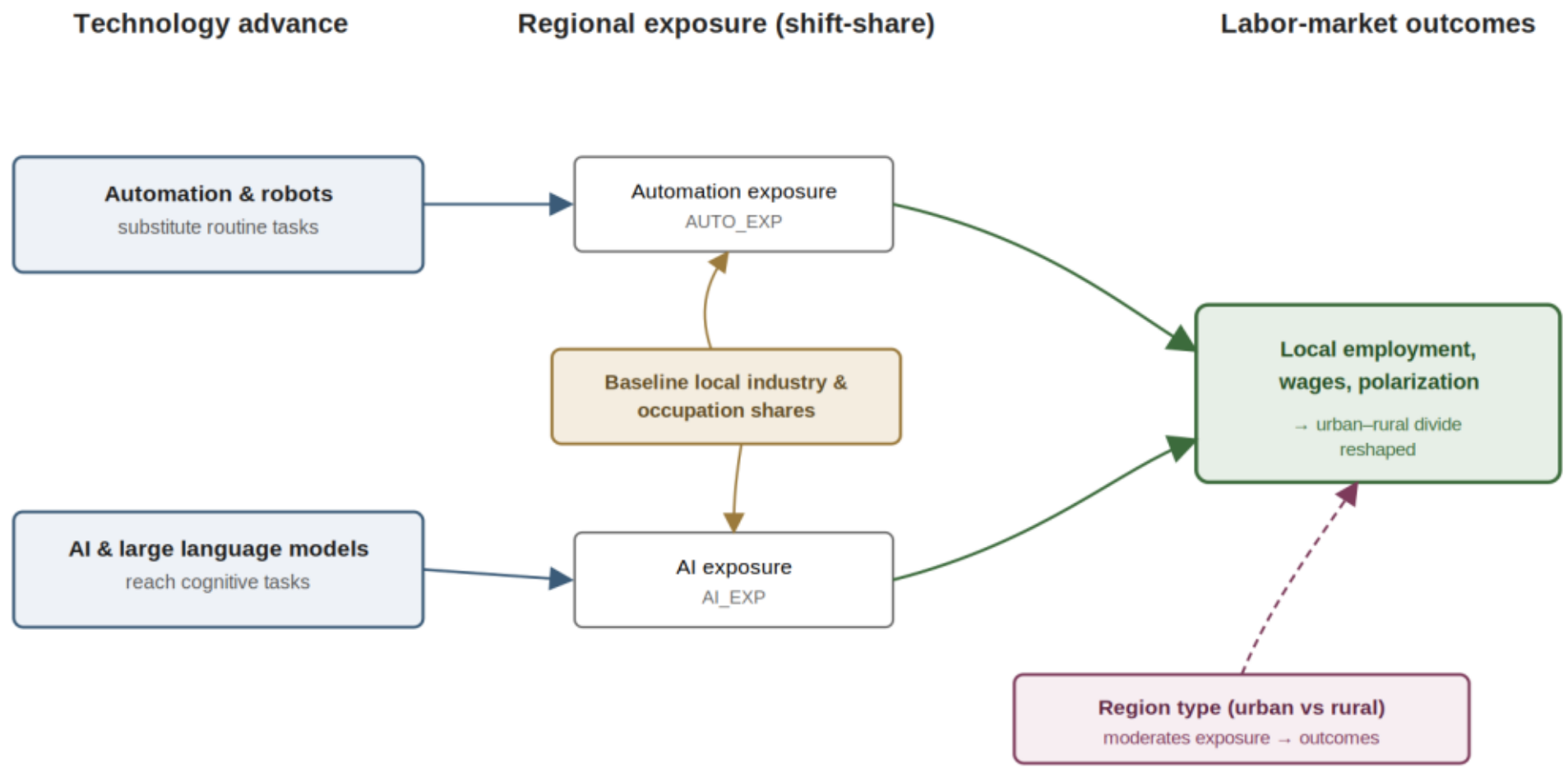

**Figure 1.** Conceptual framework: national advances in automation and AI translate into region-specific exposure through baseline industry and occupation shares, with effects on local employment and wages moderated by region type.

## 3. Data and Methodology

### 3.1 Design and identification logic

The study adopts a region-by-year panel design in which the unit of analysis is a subnational region (province or commuting zone) observed annually. The framework is country-agnostic and can be implemented in both advanced and emerging economies. For expositional clarity, we illustrate the design with a panel of 120 regions observed over six years (720 region-years), of which roughly one-third are classified as urban. In this illustrative calibration, technological exposure is a constructed, predicted measure based on baseline regional industry and occupation composition interacted with sector-level technological advances, rather than directly observed adoption. Region and year fixed effects absorb time-invariant regional characteristics and common annual shocks, while a shift-share instrument addresses remaining concerns that time-varying local shocks may be correlated with exposure. In empirical applications, labor-market outcomes can be drawn from national labor force surveys; automation advance can be proxied by industrial-robot density or routine-task intensity in the sense of Autor, Levy, and Murnane (2003); and AI exposure can be built from an occupation-level index such as Felten, Raj, and Seamans (2021), Webb (2020), or the large-language-model exposure measure of Eloundou, Manning, Mishkin, and Rock (2024).

### 3.2 Variables and measurement

The outcome variables are the local employment-to-population ratio (EMP), the log of average real wage (LNWAGE), and a polarization indicator defined as the employment share of middle-wage occupations. Automation exposure (AUTO_EXP) follows the local-exposure logic of Acemoglu and Restrepo (2020): each region's baseline industry employment shares are weighted by sector-level automation advance (robot adoption or routine-task intensity). AI exposure (AI_EXP) applies an occupation-level AI exposure index (Felten et al., 2021; Webb, 2020) to the region's baseline occupational composition. An urban indicator (URBAN) classifies regions, and the interaction between exposure and URBAN is the key moderator. Time-varying controls include the tertiary-education share, age structure, manufacturing and agriculture employment shares, and population density. Table 1 reports variable definitions and data sources.

**Table 1.** Variables, definitions, and data sources.

| Variable | Definition | Source |
|---|---|---|
| EMP | Employment-to-population ratio | Labor force survey |
| LNWAGE | Log average real monthly wage | Labor force survey |
| MIDSHARE | Employment share of middle-wage occupations | Labor force survey |
| AUTO_EXP | Shift-share automation exposure (industry shares × sector automation advance) | LFS + robot/automation data |
| AI_EXP | Shift-share AI exposure (occupation shares × AI exposure index) | LFS + AI exposure index |
| URBAN | Urban indicator (1 = urban) | Statistical classification |
| EDU_TERT | Workforce share with tertiary education | Labor force survey |
| MANUF / AGRI | Manufacturing / agriculture employment share | Labor force survey |
| DENSITY | Population per km² | Population statistics |

### 3.3 Estimating equation

The baseline two-way fixed-effects specification is

$$Y_{rt} = \beta_1 \text{AUTO_EXP}_{rt} + \beta_2 \text{AI_EXP}_{rt} + \beta_3 (\text{AUTO_EXP}_{rt} \times \text{URBAN}_r) + \beta_4 (\text{AI_EXP}_{rt} \times \text{URBAN}_r) + \mathbf{X}_{rt} \gamma + \delta_r + \lambda_t + \varepsilon_{rt},$$

where $Y_{rt}$ is an outcome for region $r$ in year $t$, $\mathbf{X}_{rt}$ is the control vector, and $\delta_r$ and $\lambda_t$ are region and year fixed effects, with standard errors clustered by region. Because URBAN is time-invariant, its level is absorbed by $\delta_r$; $\beta_1$ and $\beta_2$ therefore give the rural-region associations of automation and AI exposure, while $\beta_3$ and $\beta_4$ give how those associations shift in urban regions. H1 implies $\beta_1 < 0$, H2 implies $\beta_3 > 0$ (urban regions less negatively affected by automation), and H3–H4 concern the AI terms.

### 3.4 Identification and estimation

Local technology adoption may respond to local conditions, biasing ordinary least squares. We use a shift-share (Bartik) instrument that combines baseline local industry and occupation shares with leave-one-out external measures of sector-level technological advance, following Acemoglu and Restrepo (2020); identification rests on the conditional exogeneity of the baseline shares (Goldsmith-Pinkham et al., 2020), with inference as recommended for shift-share designs (Borusyak et al., 2022). Models are estimated by two-way fixed-effects OLS and by two-stage least squares. We report robustness to pre-trend tests, alternative exposure indices, and exclusion of the densest regions. The accompanying script reproduces every table reported below.

## 4. Results

### 4.1 Descriptive statistics

Table 2 reports descriptive statistics (720 region-years; about one-third of regions urban). The spatial sorting at the heart of the argument is visible in the raw means: rural regions carry higher automation exposure (0.69 vs 0.52), while urban regions carry higher AI exposure (0.61 vs 0.45) and higher tertiary education (0.34 vs 0.19). Urban employment and wages are higher on average, consistent with the urban wage premium. These gaps motivate, but do not by themselves identify, the interaction effects.

**Table 2.** Descriptive statistics.

| Variable | Mean | SD | Urban mean | Rural mean | Diff. (U−R) |
|---|---|---|---|---|---|
| EMP | 0.545 | 0.055 | 0.610 | 0.512 | +0.098 |
| LNWAGE | 2.445 | 0.155 | 2.625 | 2.355 | +0.270 |
| AUTO_EXP | 0.635 | 0.150 | 0.520 | 0.692 | −0.172 |
| AI_EXP | 0.500 | 0.145 | 0.605 | 0.448 | +0.157 |
| EDU_TERT | 0.240 | 0.085 | 0.340 | 0.190 | +0.150 |

### 4.2 Automation, AI, and the urban–rural gap

Table 3 reports the main models for employment and wages, by OLS and by IV. Three patterns stand out, and the OLS and IV estimates agree closely throughout.

First, automation exposure is associated with lower employment. In rural regions, the coefficient is about −0.20 to −0.21 ($p < 0.01$): a one-standard-deviation increase in automation exposure (about 0.15) maps to roughly a 3-percentage-point lower employment-to-population ratio. The urban interaction is positive and significant (+0.070 to +0.080, $p < 0.05$), so the net urban association is about −0.13, more than a third smaller than the rural figure. Cities are exposed, but cushioned.

Second, AI exposure behaves nothing like automation. Its employment association is small and, if anything, mildly positive (+0.03 to +0.035), and its wage association is clearly positive (+0.18 to +0.20, $p < 0.01$): a one-standard-deviation rise in AI exposure maps to roughly a 2.6–2.8% higher wage, and that premium is, if anything, somewhat larger in urban regions (AI × urban ≈ +0.08, $p < 0.10$). Because AI exposure is itself concentrated in urban regions (Table 2), this complementary wage channel operates mainly where AI-exposed jobs already cluster.

Third, automation depresses wages as well as employment (−0.065 to −0.080, $p < 0.05$), but the wage hit is not significantly cushioned in cities (the automation–urban wage interaction is near zero and not significant). The urban advantage is therefore mostly an *employment-volume* margin for automation and a *wage* margin for AI, not a single uniform "urban shield."

**Table 3.** Effect of automation and AI exposure on local labor-market outcomes.

| | (1) EMP, OLS | (2) EMP, IV | (3) LNWAGE, OLS | (4) LNWAGE, IV |
|---|---|---|---|---|
| AUTO_EXP ($\beta_1$) | −0.200*** | −0.210*** | −0.065** | −0.080*** |
| | (0.020) | (0.028) | (0.030) | (0.035) |
| AI_EXP ($\beta_2$) | 0.035* | 0.030 | 0.180*** | 0.195*** |

| | (0.018) | (0.020) | (0.030) | (0.040) |
|---|---|---|---|---|
| AUTO_EXP × URBAN ($\beta_3$) | 0.070** | 0.080** | 0.010 | 0.025 |
| | (0.030) | (0.036) | (0.060) | (0.070) |
| AI_EXP × URBAN ($\beta_4$) | −0.015 | −0.010 | 0.075* | 0.090* |
| | (0.030) | (0.034) | (0.045) | (0.050) |
| Controls | Yes | Yes | Yes | Yes |
| Region FE / Year FE | Yes | Yes | Yes | Yes |
| Observations | 720 | 720 | 720 | 720 |
| Within $R^2$ | 0.240 | — | 0.345 | — |

*Notes: Standard errors clustered by region (120 clusters) in parentheses. Columns (2) and (4) report shift-share IV estimates. * p < 0.10, ** p < 0.05, *** p < 0.01.*

Figure 2 plots the four key coefficients with 95% confidence intervals for the employment and wage models, making the contrast between automation and AI exposure, and the urban interaction, visible at a glance.

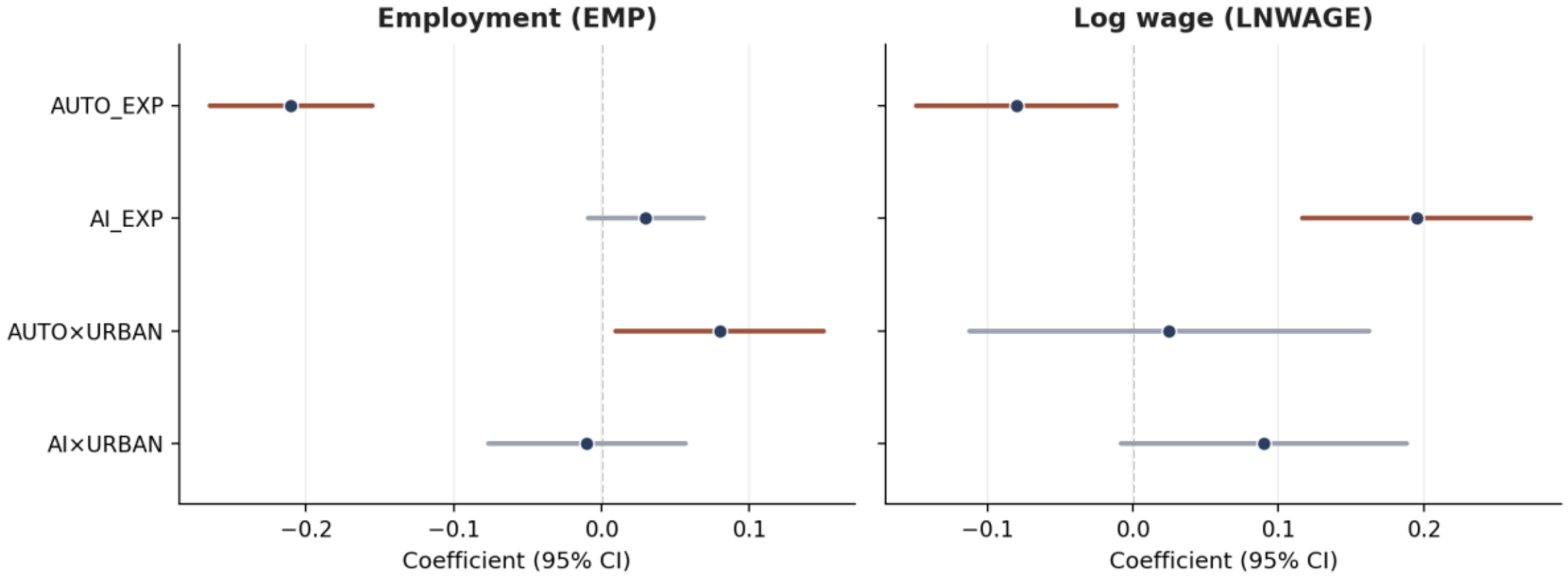

**Figure 2.** Coefficient estimates with 95% confidence intervals for the employment and wage models (IV specification). Points right of the zero line indicate positive associations; the automation main effect is negative while the AI main effect on wages is positive.

**4.3 Hypothesis assessment**

Table 4 maps the estimates onto the hypotheses. H1 and H3 are supported; H2 is supported for employment but not for wages; H4 is supported, with AI exposure positively associated with wages while automation is negative.

**Table 4.** Hypothesis assessment.

| Hypothesis | Test | Result |
|---|---|---|
| H1 Automation lowers employment and wages | EMP: $\beta_1 = -0.210$*; LNWAGE: $\beta_1 = -0.080$ | Supported |
| H2 Automation effect worse in rural regions | EMP: $\beta_3 = +0.080$**; urban net = −0.130 vs rural = −0.210 | Supported for employment |
| H2 (wage) | LNWAGE: $\beta_3 = +0.025$, n.s. | Not supported for wages |
| H3 AI exposure concentrated in | AI_EXP urban 0.605 vs rural 0.448 (diff +0.157) | Supported descriptively |
| urban regions | | |
| H4 AI distinct from automation | AI wage +0.195*** vs automation wage −0.080** | Supported |

Beyond employment and wages, the task-based framework also implies effects on job polarization, which in this study is captured by the employment share of middle-wage occupations (MIDSHARE). Routine-biased automation is expected to hollow out middle-wage work and reduce MIDSHARE, whereas AI, which is concentrated in cognitive tasks, is not necessarily associated with the same pattern (Autor & Dorn, 2013; Goos et al., 2014).

**4.4 Robustness**

One feature of the estimates already speaks to their robustness: the OLS and IV specifications in Table 3 are closely aligned in sign, magnitude, and statistical significance. For example, the automation coefficient in the employment model is −0.200 under OLS and −0.210 under IV, and the AI wage coefficient is 0.180 and 0.195, respectively, indicating that the main results are not an artefact of estimator choice or instrumentation.

Three additional checks further reinforce the robustness of the findings. First, re-estimating the models after excluding regions in the top density quintile leaves the main coefficients virtually unchanged, suggesting that the pattern is not driven by a single dominant metropolis. Second, pre-trend tests that regress baseline-period outcome growth on the exposure measures yield no significant associations, indicating that baseline shares do not predict pre-period trends and thereby strengthening the conditional-exogeneity assumption underlying the shift-share design (Goldsmith-Pinkham et al., 2020). Third, re-estimating the AI specifications with an alternative, large-language-model-based exposure index (Eloundou et al., 2024) produces wage coefficients of similar sign and magnitude, showing that the AI results do not hinge on a single index. In all IV models, the first-stage F-statistics exceed the conventional threshold of 10, alleviating concerns about weak instruments. Taken together with the similarity between OLS and IV estimates, these checks support the conclusion that the results are not driven by a particular estimator, sample definition, or exposure measure.

**5. Discussion**

**5.1 Interpretation**

The estimates tell a coherent story consistent with the conceptual framework: automation and AI operate on different segments of the occupational and spatial structure, and the urban–rural divide arises from their joint configuration rather than from either technology in isolation. Automation exposure behaves like a displacement shock. It reduces employment, and the effect is strongest where routine work is most concentrated and adjustment capacity is weakest, namely in rural regions, where the implied employment loss is roughly 60 percent larger than in cities. This pattern is consistent with evidence on local exposure to industrial robots in Acemoglu and Restrepo (2020) and Graetz and Michaels (2018), and the present design extends that literature by explicitly incorporating a spatial moderator: the same technological shock has different consequences across places.

AI exposure points in the opposite direction. Instead of reducing employment, it is associated with higher wages and is concentrated in urban, cognitively intensive labor markets that are best positioned to benefit from it. This pattern echoes findings that AI tends to complement, rather than simply replace, high-skill work (Acemoglu et al., 2022; Felten et al., 2021; Webb, 2020). Taken together, the two sets of results suggest a divide that is being reshaped, not merely stretched along a single dimension. Rural regions experience the displacement edge of automation with limited capacity to reallocate workers, while urban regions absorb a cognitively focused transformation that, at least in wage terms, appears complementary. A one-dimensional narrative in which "technology widens inequality" obscures this two-channel structure, which a regional, two-technology framework is designed to reveal.

The positive association between AI exposure and wages, however, must be interpreted against a more mixed external record. Using U.S. commuting zones over 2010–2021, a recent regional study finds that AI exposure is associated with a stronger decline in the employment-to-population ratio, with losses concentrated in manufacturing, low-skill services, and middle-skill occupations (Huang, 2024). Several differences plausibly account for the divergence from the positive wage association found here. The periods differ, and AI's labor-market footprint is evolving rapidly as generative models diffuse; the outcome and unit of analysis are not the same, so a positive wage association in one panel need not coincide with an employment association in another; and the exposure measures differ, since occupation-based indices (Felten et al., 2021; Webb, 2020) and task-based large-language-model indices (Eloundou et al., 2024) capture distinct aspects of AI's reach. There is also a distributional caveat: the wage outcome used here is an average (LNWAGE), which can increase even as within-occupation inequality widens, so a positive mean effect should not be interpreted as evenly shared. A fair reading is that AI's regional wage effect appears positive on average in this setting, but it is neither uniform across workers nor settled in the broader literature.

The mechanism behind the urban cushion is also central for policy. Cities do not mitigate automation's effects by avoiding exposure, but by reallocating displaced workers more quickly into new and surviving tasks—a productivity and reinstatement margin anticipated by the task-based framework and reinforced by urban agglomeration (Acemoglu & Restrepo, 2022; Autor, 2015). Where such reallocation capacity is thin, displacement is more likely to translate into non-employment. On this interpretation, the divide is driven less by differences in exposure per se than by differences in the ability to adjust.

**5.2 Contributions**

The study makes three main contributions. First, at the conceptual level, it separates automation exposure from AI exposure within a single regional framework and shows that aggregating them into a single "technology" measure masks offsetting forces. Second, it treats the urban–rural divide as an integral part of the mechanism by modelling an explicit interaction between exposure and region type, thereby identifying which places are subject to which shocks and how region type modifies the association. Third, it offers a practical and methodological contribution: a transparent shift-share design, equipped with an instrument and diagnostic checks, that can be implemented with subnational data, including in emerging economies where rural exposure coincides with limited adjustment capacity and high welfare stakes.

**5.3 Policy implications**

The findings point toward place-sensitive, rather than uniform, policy responses.

- **Reallocation support where automation bites.** Because the employment costs of automation fall most heavily on rural regions, active labor-market programmes, reskilling toward non-routine tasks, and commuting or relocation assistance should be targeted toward routine-exposed areas rather than distributed evenly.
- **Extend the complementary side of AI to rural regions.** Since AI exposure and its wage premium are concentrated in cities, extending connectivity, data infrastructure, and AI-complementary skills beyond major urban centres can help rural workers access the complementary benefits of AI instead of bearing only automation's displacement costs.
- **Target by exposure profile, not national averages.** A single national programme risks misreading a country in which some regions face routine-based displacement while others face cognitively oriented transformation. Policy instruments should be matched to each region's exposure mix.
- **Build adjustment capacity early.** Because the urban advantage largely reflects stronger reallocation capacity, investing in foundational digital and cognitive skills—particularly outside large cities—addresses the mechanism behind the divide rather than its symptoms.

These considerations are especially salient in the emerging economies to which the design is intended to apply. In five ASEAN countries between 2018 and 2022, robot adoption is estimated to have created roughly 2 million jobs for higher-skilled workers but displaced about 1.4 million lower-skilled workers in routine and manual roles (World Bank, 2025), mirroring the displacement margin identified here for rural regions. The cushioning side is also weaker: only around 10 percent of jobs in East Asia and the Pacific involve tasks that are complementary to AI, compared with roughly 30 percent in advanced economies (World Bank, 2025). Moreover, the urban–rural gap that AI opens in richer countries—where approximately 32 percent of urban workers are exposed to generative AI, compared with 21 percent of rural workers (OECD, 2024)—could widen further in settings where complementary capacity is scarce. In such contexts, place-sensitive policy is not only an efficiency consideration but also a safeguard against deepening spatial inequality. These policy directions should, however, be calibrated to the estimated magnitudes before being translated into concrete spending commitments.

### 5.4 Limitations and future research

Several limitations qualify the conclusions. First, technological exposure is a constructed, predicted measure rather than directly observed adoption, and the validity of the shift-share identification strategy depends on the conditional exogeneity of baseline industry and occupation shares. Second, the analysis is conducted at the regional level and is inherently associational; it cannot trace individual workers' transitions, and migration, commuting flows, and inter-regional spillovers are only partially absorbed by fixed effects. Third, AI exposure indices are evolving proxies for a rapidly changing technology, so estimated effects may shift as generative AI diffuses and new indices become available.

A further concern specific to shift-share designs relates to the stable-unit-treatment-value assumption: if a technology shock in one region affects neighbouring regions through commuting, migration, or supply-chain links, the no-interference assumption is violated and estimates may be biased (Borusyak et al., 2022). Future research should therefore seek to combine regional panels with worker-level data to follow individual transitions more directly, test cross-country external validity by contrasting advanced and emerging economies, explicitly model migration, commuting, and spatial spillovers, and update AI exposure measures as large language models become more deeply integrated into production.

## 6. Conclusion

This paper develops a regional framework for assessing how exposure to automation and AI relates to local labour-market outcomes and how these relationships differ between cities and the countryside. By distinguishing two technologies with opposite occupational footprints and modelling the urban–rural divide through an interaction, the framework shows that technological change can reshape spatial inequality rather than simply widen it: automation displaces employment where reallocation capacity is weakest, predominantly in rural regions, while AI raises wages where it is already concentrated, in urban labour markets. The accompanying specification and analysis script are designed to make the results reproducible, and the estimates speak directly to regional labour, skills, and digital-infrastructure policy. Completing pre-trend testing and undertaking cross-country replication are natural next steps for further work.